\documentclass[twocolumn,showpacs,preprintnumbers,amsmath,amssymb]{revtex4}
\usepackage{graphicx}
\usepackage{dcolumn}
\usepackage{bm}
\begin{document}       

\title{Aeolian transport layer}
\author{Murilo P. Almeida}          
\author{Jos\'e S. Andrade Jr.}
\author{Hans J. Herrmann}
\altaffiliation[Also at ] {Institute for Computer Physics,\\
University of Stuttgart\\ email: hans@ica1.uni-stuttgart.de }
\affiliation{Departamento de F\'{\i}sica, Universidade Federal do Cear\'a,\\ 
60455-900 Fortaleza, Cear\'a, Brazil}

\date{\today}

\begin{abstract}
We investigate the airborne transport of particles on a granular
surface by the saltation mechanism through numerical simulation of
particle motion coupled with turbulent flow. We determine the
saturated flux $q_{s}$ and show that its behavior is consistent with a
classical empirical relation obtained from wind tunnel
measurements. Our results also allow to propose a new relation valid
for small fluxes, namely, $q_{s}=a(u_{*}-u_{t})^{\alpha}$, where
$u_{*}$ and $u_{t}$ are the shear and threshold velocities of the
wind, respectively, and the scaling exponent is $\alpha \approx 2$. We
obtain an expression for the velocity profile of the wind distorted by
the particle motion and present a dynamical scaling relation. We also
find a novel expression for the dependence of the height of the
saltation layer as function of the wind velocity.
\end{abstract}


\maketitle

The transport of sand by wind is among others responsible for sand
encroachment, dune motion and the formation of coastal and desert
landscapes. The dominating transport mechanism is saltation as first
described by Bagnold \cite{Bagnold44} which consists of grains being
ejected upwards, accelerated by the wind and finally impacting onto
the ground producing a splash of new ejected particles. Reviews are
given in Refs.~\cite{Anderson91,Herrmann04}. Quantitatively
this process is however far from being understood.

Due to Newton's second law the wind loses more momentum with
increasing number of airborne particles until a saturation is
reached. The maximum number of grains a wind of given strength can
carry through a unit area per unit time defines the saturated flux of
sand $q_s$. This quantity has been measured by many authors in wind
tunnel experiments and on the field, and numerous empirical expressions
for its dependence on the strength of the wind have been proposed
\cite{Lettau78,Werner90,Rasmussen91,Zhou02}. In previous studies
theoretical forms have also been derived using approximations for the
drag in turbulent flow \cite{Owen64,Sorensen85}. All these relations
are expressed as polynomials in the wind shear velocity $u_{*}$ which are of
third order, under the assumption that the grain hopping length scales
with $u_{*}$ \cite{Bagnold44,Lettau78,Owen64,Sorensen85,White79} and
otherwise can be more complex \cite{Werner90}. The velocity profile in
a particle laden layer has also been the object of measurement
\cite{Butterfield93,Nishimura00} and modellization~\cite{Ungar87}. 
Surprisingly however very few  measurements of the height of the
saltation layers as function of $u_{*}$ have been reported \cite{Dong02}
and no systematic data close to the threshold are available. The
complete analytical treatment of this problem remains out of reach not
only because of the turbulent character of the wind, but also due to
the underlying moving boundary conditions in the equations of
motion. Despite much research in the past \cite{Nalpanis93} there
remain many uncertainties about the trajectories of the particles and
their feedback with the velocity field of the wind. It is this
challenge which motivates the present work. 

We will present the first numerical study of saltation which solves
the turbulent wind field and its feedback with the dragged
particles. As compared to real data, our values have no experimental
fluctuations neither in the wind field nor in the particle sizes. As a
consequence, we can determine all quantities with higher precision
than ever before, and therefore with a better resolution close to the
critical velocity at which the saltation process starts.

In order to get quantitative understanding of the layer of airborne
particle transport above a granular surface, we simulate the situation
inside a two-dimensional channel with a mobile top wall as
schematically shown in Fig.~1.
\begin{figure}
\begin{center}
\includegraphics[width=8.0cm]{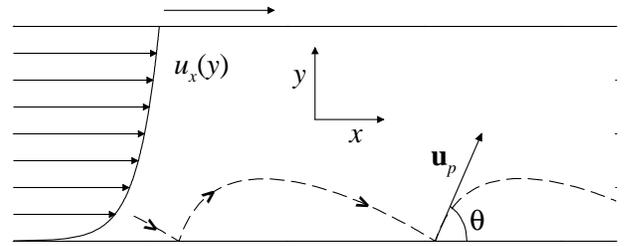} 
\end{center}
\caption{Schematic representation of the setup showing the mobile wall 
at the top, the velocity field at different positions in the {\it
y}-direction and the trajectory of a particle stream (dashed
line). At the collision between this stream and the static wall at the
bottom, we consider that the particles rebound to the air flow at a
ejection angle $\theta$, with only a fraction $r$ of their
original kinetic impact energy.}
\end{figure}
We impose a pressure gradient between the left and the right
side. Gravity points down, i.e., in negative {\it y}-direction. The
{\it y}-dependence of the pressure drop is adjusted in such a way as
to insure a logarithmic velocity profile along the entire channel in
the case without particles, as it is expected in fully developed
turbulence \cite{Prandtl}. More precisely, this profile follows the
classical form
\begin{equation}\label{Eq_logprof}
u_{x}(y)=(u_{*}/\kappa) ln({y/y_{0}})~,
\end{equation}
where $u_{x}$ is the component of the wind velocity in the {\it
x}-direction, $u_{*}$ is the shear velocity, $\kappa=0.4$ is the von
Karman constant and $y_0$ is the roughness length which is typically
between $10^{-4}$ and $10^{-2} m$. The upper wall of the channel is
moved with a velocity equal to the velocity of the wind at that
height in order to insure a non-slip boundary condition.

The fluid mechanics inside the channel is based on the assumptions
that we have an incompressible and Newtonian fluid flowing under
steady-state and homogeneous turbulent conditions. The fluid is air
with viscosity $\mu=1.7894 \times 10^{-5} kg~m^{-1}s^{-1}$ and density
$\rho=1.225~kg~m^{-3}$.  The Reynolds-averaged Navier-Stokes equations
with the standard $k-\epsilon$ model are used to describe
turbulence. The numerical solution for the velocity and pressure
fields is obtained through discretization by means of the control
volume finite-difference technique \cite{Patankar80,Fluent}. The
integral version of the governing equations is considered at each cell
of the numerical grid to generate a set of non-linear algebraic
equations which are pseudo-linearized and solved. The criteria for
convergence used in the simulations are defined in terms of residuals,
i.e., a measure of the degree to which the conservation equations are
satisfied throughout the flow field. In our simulations convergence is
achieved only when each of the normalized residuals falls below
$10^{-6}$.

After having produced a steady-state turbulent flow, we proceed with
the simulation of the particle transport along the channel. Assuming
that drag and gravity are the only relevant forces acting on the
particles, their trajectory can be obtained by integrating the
following equation of motion:
\begin{equation}\label{Eq_motion}
\frac{d{\bf u}_{p}}{dt}=F_{D}({\bf u}-{\bf u}_{p})
+{\bf g}(\rho_{p}-\rho)/\rho_{p}~,
\end{equation}  
where $u_p$ is the particle velocity, ${\bf g}$ is gravity and
$\rho_{p}=2650\ kg\ m^{-3}$ is a typical value for the density of sand
particles. The term $F_{D}({\bf u} - {\bf u}_{p})$ represents the drag
force per unit particle mass where
\begin{equation}\label{Eq_prefactor} 
F_{D}=\frac{18\mu}{\rho_p d^{2}_{p}} \;\; \frac{C_D {\rm Re}}{24}~, 
\end{equation}
$d_{p}=2.5\times 10^{-4}\ m$ is a typical particle diameter, ${\rm
Re}\equiv\rho d_{p} \left\vert {\bf u}_{p}-{\bf u} \right\vert/\mu$ is
the particle Reynolds number, and the drag coefficient $C_{D}$ is
taken from empirical relations \cite{Morsi72}. Each particle
in our calculation represents in fact a stream of real grains. It is
necessary to take into account the feedback on the local fluid
velocity due to the momentum transfer to and from the
particles. Specifically, this coupling effect is considered here by
alternately solving the discrete and continuous phase equations until
the solutions in both phases agree. The momentum transfer from one
phase to another is computed by adding the momentum change of every
particle as it passes through a control volume \cite{Fluent},
\begin{equation}\label{Eq_coupling} 
{\bf F}=\sum_{\rm particles}F_{D}({\bf u}-{\bf u}_{p})\dot{m}_{p}\Delta t 
\end{equation}
where $\dot{m}_p$ is the mass flow rate of the particles and $\Delta
t$ the time step. The exchange term Eq.~(\ref{Eq_coupling}) appears
as a sink in the continuous phase momentum balance.

In Fig.~1 we see the trajectory of one particle stream and the
velocity vectors along the {\it y}-direction.  Each time a particle
hits the ground it loses a fraction $r$ of its energy and a new
stream of particles is ejected at that position with an angle
$\theta$. The parameters $r=0.84$ and $\theta=36^\textrm{o}$ are
chosen from experimental measurements \cite{Anderson88,Rioual00}.  
We also studied other values for $r$ and $\theta$ and even considered a
continuous distribution of ejection angles. As expected, the choice of
unrealistic values produces unphysical results. More details will be
given in Ref.~\cite{Almeida05}.

If $u_{*}$ is below a threshold value $u_{t}$ the energy loss at each
impact prevails over the energy gain during the acceleration through
drag and particle transport comes to a halt as illustrated in Fig.~2.
\begin{figure}
\includegraphics[width=8.0cm]{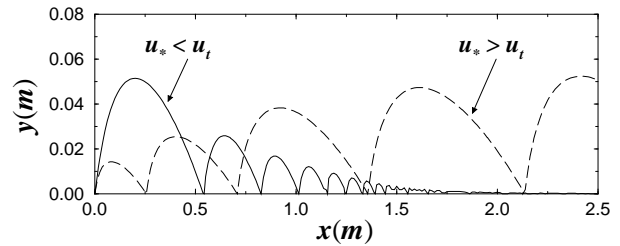}
\caption{Typical trajectories of particles computed for $u_{*}<u_{t}$ 
(full line) and $u_{*}>u_{t}$ (dashed line).}
\end{figure}
Only for $u_{*} > u_{t}$ steady sand motion is achieved. 
The resulting flux is given by
\begin{equation}\label{Eq_flux}
q=m_{p}n_{p}{\langle u_{p} \rangle}
\end{equation}
where $m_{p}$ and $\langle u_p \rangle$ are the mass and the average
velocity of the particles, respectively, and $n_{p}$ is the number of
particle streams released. The first added particle streams are
strongly accelerated in the channel and their jumping amplitude
increases after each ejection until a maximum is reached as seen in
Fig.~2. The more particles are injected the smaller is this final
amplitude. Beyond a certain number $n_{s}$ of particle streams, the
trajectories however start to lose energy and the overall flux is
reduced. This critical value $n_s$ characterizes the saturated flux
$q_s$ through Eq.~(\ref{Eq_flux}).

In Fig.~3 we see the plot of $q_s$ as function of the wind velocity
$u_*$.  Clearly, there exists a critical wind velocity threshold $u_t$
below which no sand transport occurs at all. This agrees well with
experimental observations~\cite{Lettau78,Bagnold38}. Also shown in
Fig.~3 is the best fit to the simulation data using the classical
expression proposed by Lettau and Lettau \cite{Lettau78},
\begin{equation}\label{Eq_Lettau}
q_s=C_{L}\frac{\rho}{g}u_{*}^{2}(u_{*}-u_{t})~,
\end{equation} 
where $C_{L}$ is an adjustable parameter. We find very good agreement
using fit parameters of the same order as those of the original work
and a threshold value of $u_{t}=0.35 \pm 0.02$. This is in fact, to
our knowledge, the first time a numerical calculation is able to
quantitatively reproduce this empirical expression and it confirms the
validity of our simulation procedure. Other empirical relations from
the literature \cite{Owen64,Sorensen85,White79} can also be used to
fit these results \cite{Almeida05}. Close to the critical velocity
$u_{t}$ interestingly we find that a parabolic expression of the form
\begin{equation}\label{Eq_power}
q_s = a (u_*-u_t)^2 
\end{equation} 
fits the data at least as well as Eq.~(\ref{Eq_Lettau}), as can be
seen in the inset of Fig.~3. The threshold obtained in this case,
namely, $u_{t}=0.33 \pm 0.01$, is slightly lower than the one obtained
for the classical cubic expression Eq.~(\ref{Eq_Lettau}). We believe
that our parabolic expression describes correctly the critical behavior
very close the transition right above $u_t$, since at this point the 
classical assumption of proportionality between the saltation jump 
length and $u_{*}$ cannot hold.
\begin{figure}
\includegraphics[width=8.0cm]{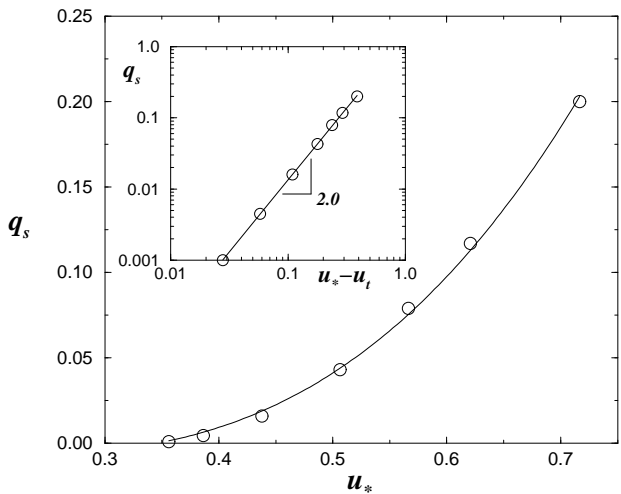}
\caption{Saturated flux $q_{s}$ as function of $u_{*}$. The full line is 
the fit using the expression proposed by Lettau and Lettau
\cite{Lettau78}, $q_s \propto u_{*}^{2}(u_{*}-u_{t})$, with 
$u_t=0.35 \pm 0.02$. The inset shows the same data in a double
logarithmic plot. The straight line corresponds to the least-squares
fit to the data of the power-law relation, $q_s \propto
(u_*-u_t)^\alpha$, with the exponent $\alpha=2.01 \pm 0.01$ and the
critical point given by $u_t=0.33 \pm 0.01$.}
\end{figure}

Experimentally much more difficult to access is the velocity profile
of the wind within the layer of grain transport. This profile clearly
deviates from the undisturbed logarithmic form of
Eq.~(\ref{Eq_logprof}) because of the momentum the fluid must locally
exchange with the particles. In Fig.~4 we show the loss of velocity with
respect to the logarithmic profile without particles of
Eq.~(\ref{Eq_logprof}) for different values of $q$ as function of the
height $y$.
\begin{figure}
\includegraphics[width=8.0cm]{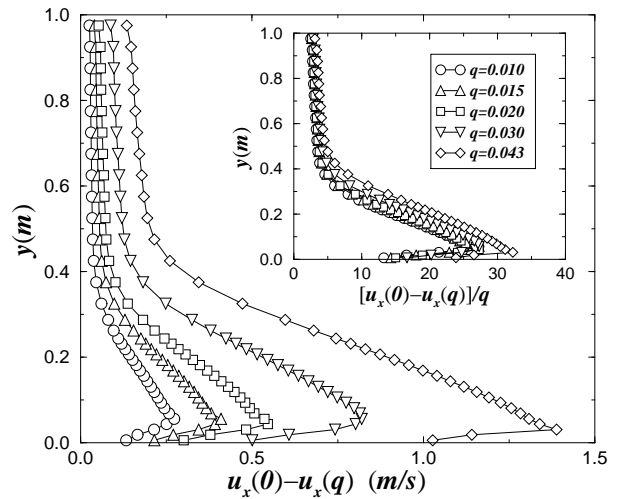}
\caption{Profile of the velocity difference $u_{x}(0)-u_{x}(q)$ for 
different values of the flux $q$ at $u_{*}=0.51$. The inset shows the 
data collapse of these data obtained by rescaling the velocity difference
with the corresponding $q$.}
\end{figure}
As clearly seen in Fig.~4, the loss of velocity is maximal at the same
height $y_{max}$, regardless of the value of flux $q$. Except for
large values of the flux, dividing the velocity axis by $q$ one can
collapse all the profiles quite well on top of each other as can be
verified in the inset of Fig.~4.
\begin{figure}
\includegraphics[width=8.0cm]{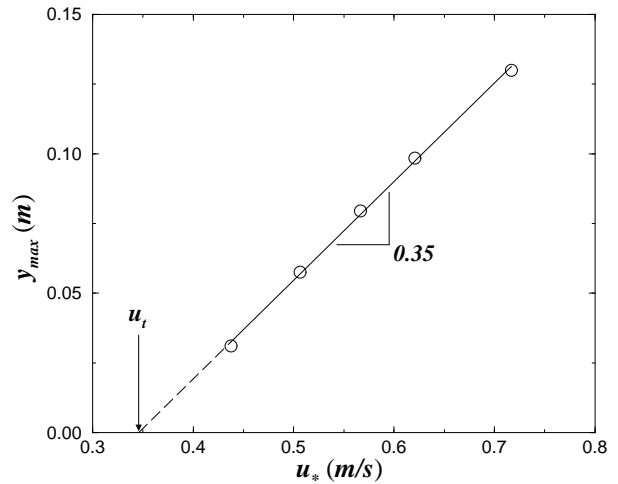}
\caption{Height $y_{max}$ of the maximum loss of velocity
as function of $u_*$. The height $y_{s}$ of the largest probability to
find a particle coincides with $y_{max}$. The solid line corresponds
to the best linear fit to the data with a slope equal to 0.35. By
extrapolation, the intercept with the {\it x}-axis provides an
alternative estimate for the critical point, $u_{t}=0.35~m/s$, that is
consistent with the other calculations.}
\end{figure}

The position $y_{max}$ of the height of maximum loss depends
essentially linearly on $u_*$ as shown in Fig.~5. This is consistent
with the observation that the saltation jump length is proportional to
$u_{*}$ \cite{Sorensen85}. The proportionality constant obtained from
the best linear fit to the data is $0.35~s$. Quantitatively the data
in Fig.~5 also fit very well into the experiment data plots of
Ref.~\cite{Dong02}. By extrapolation to $y_{max}=0$, we obtain an
alternative estimate for the threshold velocity, $u_{t}=0.35~m/s$,
that is consistent with the values calculated before from the fits to
the data using Eqs.~(\ref{Eq_Lettau}) and (\ref{Eq_power}).

Whoever has been in the desert or on a beach during a very windy day
knows that the saltation process in nature looks like a sheet of
particles floating above ground at a certain height $y_{s}$ which
strongly depends on the wind velocity. This height corresponds to the
position of the largest likelihood to find a particle as obtained from
the maximum of the density profile of particles as function of height
$y$. Fig.~5 implies that the profile of velocity difference of the
wind has a minimum at a similar height, which is consistent with the
maximal loss of momentum. Within the error bars our results in fact
yield that $y_{s}$ coincides with the values of $y_{max}$ in
Fig.~5. More details will be presented in Ref.~\cite{Almeida05}. It is
important to note that both heights, $y_{max}$ and $y_s$, also have the
same linear dependence on $u_*$.

We have shown in this letter results of simulations giving insight
about the layer of granular transport under conditions of turbulent
flow. The lack of experimental noise allows for a precise study close
to the critical threshold velocity $u_t$ that lead us to a parabolic
dependence of the saturated flux. In addition, we show that the
velocity profile disturbed by the presence of grains scales linearly
with the flux of grains, except close to saturation. Notably a
characteristic height appears at which the momentum loss in the fluid
and the grain density are maximized. Moreover, this height increases
linearly with the wind velocity $u_*$. It would be very interesting to
verify experimentally these novel predictions. The present model can
be extended in many ways including the study of the dependence of the
aeolian transport layer on the grain diameter, the gas viscosity, 
and the solid or fluid densities. This would allow to calculate,
for instance, the granular transport on Mars and compare with the
expression presented in Ref.~\cite{White79}. Work in this direction is
under way \cite{Almeida05}.

We thank Keld Rasmussen, Eric Parteli, Josu\'{e} Mendes Filho and
Asc\^{a}nio Dias Ara\'{u}jo for discussions and CNPq, CAPES, FUNCAP
and the Max-Planck prize for financial support.

\bibliographystyle{prsty}

\begin{thebibliography}{10}

\bibitem{Bagnold44} 
R. A. Bagnold, {\it The Physics of blown sand and desert dunes}
(Methuen, London, 1941).

\bibitem{Anderson91} 
R. S. Anderson, M. S{\o}rensen and B. B. Willetts, Acta Mech. (Suppl.)
{\bf 1}, 1 (1991).

\bibitem{Herrmann04}
H. J. Herrmann, in {\it The Physics of Granular Media},
eds. H. Hinrichsen and D. Wolf (Wiley VCH, Weinheim, 2004),
p. 233-252.

\bibitem{Lettau78}
K. Lettau and H. Lettau, in {\it Exploring the world's driest
climate}, eds. H. Lettau and K. Lettau, Center for Climatic Research
(Univ. of Wisconsin, Madison, 1978).

\bibitem{Werner90}
B. T. Werner, J. Geol. {\bf 98}, 1 (1990).

\bibitem{Rasmussen91} 
K. R. Rasmussen and H. E. Mikkelsen, Acta Mechanica Suppl. 1, {\bf
135} (1991).

\bibitem{Zhou02}
Y.-H. Zhou, X. Guo and X. J. Zheng, Phys. Rev. E {\bf 66}, 021305
(2002).

\bibitem{Owen64}
P. R. Owen, J. Fluid Mech. {\bf 20}, 225 (1964).

\bibitem{Sorensen85} 
M. S{\o}rensen, in {\it Proc. Int. Wkshp. Physics of Blown Sand}, Vol.1
(Univ. of Aarhus, Denmark, 1985), p. 141; Acta Mech. (Suppl.) {\bf
1}, 67 (1991).

\bibitem{White79} 
B. R. White, Geophys. Res. {\bf 84}, 4643 (1979).

\bibitem{Butterfield93} 
G. R. Butterfield, in {\it Turbulence: Perspectives on Flow and
Sediment Transport}, eds. N. J. Cliffod, J. R. French and J. Hardisty,
Chapter 13 (John Wiley, 1993), p.305.

\bibitem{Nishimura00} 
K. Nishimura and J. C. R. Hunt, J. Fluid. Mech.  {\bf 417}, 77 (2000).

\bibitem{Ungar87} 
J. E. Ungar and P. K. Haff, Sedimentology {\bf 34}, 289 (1987).

\bibitem{Dong02}
Z. Dong, X. Liu, H. Wang, A. Zhao and X. Wang, Geomorphology {bf 49},
219 (2002).

\bibitem{Nalpanis93} 
P. Nalpanis, J. C. R. Hunt and C. F. Barrett, J. Fluid Mech. {\bf
251}, 661-685 (1993).

\bibitem{Prandtl}
L. Prandtl, in {\it Aerodynamic Theory}, ed. W. F. Durand, Vol. III
(Springer, Berlin, 1935), p.34.

\bibitem{Patankar80}
S. V. Patankar, \textit{Numerical Heat Transfer and Fluid Flow} 
(Hemisphere, Washington DC, 1980)

\bibitem{Fluent}
The {FLUENT} (trademark of {FLUENT} Inc.)  commercial package for
fluid dynamics analysis is used in this study.

\bibitem{Morsi72}
S. A. Morsi and A. J. Alexander, J. Fluid Mech. {\bf 55}, 193 (1972).

\bibitem{Anderson88} 
R. S. Anderson and P. K. Haff, Science {\bf 241}, 820 (1988).

\bibitem{Rioual00} 
F. Rioual, A. Valance and D. Bideau, Phys. Rev. E {\bf 62}, 2450-2459
(2000).

\bibitem{Almeida05} M. P. Almeida, J. S. Andrade Jr. and H. J. Herrmann, 
preprint in preparation.

\bibitem{Bagnold38}
R. A. Bagnold, Proc. R. Soc. Lond A {\bf 167}, 282 (1938).

\end{thebibliography}

\end{document}